\documentclass[a4paper,12pt]{article}
\usepackage{amsmath,amsfonts,amssymb}
\usepackage{cite}

\begin{document}

\renewcommand*{\thefootnote}{\fnsymbol{footnote}}

\begin{center}
{\Large\bf The experimental observation of $a_0(1710)$: Long awaited from Regge approach}
\end{center}

\begin{center}
{S. S. Afonin\footnote{E-mail: \texttt{s.afonin@spbu.ru}}}
\end{center}

\begin{center}
{\small Saint Petersburg State University, Universitetskaya nab. 7/9, St. Petersburg, 199034, Russia}\\
{\small 
NRC "Kurchatov Institute": Petersburg Nuclear Physics Institute, mkr. Orlova roshcha 1, Gatchina, 188300}
\end{center}

\renewcommand*{\thefootnote}{\arabic{footnote}}
\setcounter{footnote}{0}

\begin{abstract}
Recently, the BABAR (in 2021), BESIII (in 2022), and LHCb (in 2023) Collaborations reported the observation of the $a_0(1710)$ resonance. This
has sparked a lively debate in the literature about the nature of this possible isospin partner of $f_0(1710)$, since the latter has long been regarded as
the main candidate for the lightest glueball.
We highlight the clear prediction of $a_0(1710)$ in 2007 within the Regge approach using the observed hydrogen-like degeneracy in the spectrum of light mesons.
Our reanalysis of the data used shows that the prediction was reliable and thus indicates that $a_0(1710)$ and $f_0(1710)$ are conventional quark-antiquark states.
\end{abstract}

\bigskip

Recently (in 2021-2023) the observation of $a_0(1710)$-meson has been reported by the BABAR~\cite{BaBar:2021fkz}, BESIII~\cite{BESIII:2022npc}, and LHCb~\cite{LHCb:2023evz}
Collaborations in the decays of heavy $\eta_c$, $D_s$, and $B$ mesons with $\eta\pi\pi$ or $KK\pi$ in the final states. The averaged value of its mass in the modern Particle
Data~\cite{pdg} is
\begin{equation}
\label{mass}
M_{a_0(1710)}=1713\pm19\,\text{MeV}.
\end{equation}
The nature of this resonance is a subject of debate, see~\cite{Guo:2022xqu,Achasov:2023izs,Ding:2024lqk,Song:2025tog} and references therein.
The value of its mass suggests that $a_0(1710)$ represents the isovector partner of $f_0(1710)$~\cite{pdg}. This feature makes the existence of $a_0(1710)$, if confirmed,
especially intriguing: It has long been thought that $f_0(1710)$ has a large glueball component (see, e.g.,~\cite{Janowski:2014ppa} and references  therein
and also the related discussions in~\cite{Guo:2022xqu}) and for this reason $f_0(1710)$ should not have any isospin partner.

The $a_0$-meson with mass close to~\eqref{mass} was theoretically predicted within the framework of various approaches, a brief summary of these predictions is given in Table~1.
Perhaps the first prediction was made within the Godfrey--Isgur relativized potential model~\cite{Godfrey:1985xj} in 1985.
In terms of nonrelativistic $n^{2S+1}\!L_{J}$ basis, it can be identified with the state $2^3\!P_0$. From the modern perspective, the description of radial excitations of light mesons
given by the old Godfrey--Isgur model is not accurate. For instance, this model yields no prediction for $a_0(1450)$ (see Fig.~3 in~\cite{Godfrey:1985xj}). A more accurate
description of radial spectrum of light mesons is given by the radial Regge trajectories~\cite{ani,bugg}. To the best of our knowledge, the first clear prediction of $a_0(1710)$
mass and total decay width within the Regge approach was made in Ref.~\cite{cl2} using earlier analysis of Refs.~\cite{cl1b,cl2b}. Since this point is somehow overlooked in the literature when discussing the nature of $a_0(1710)$, we think it useful to remind the reader how the prediction of this meson followed from the mentioned Regge analysis.

\begin{table}

  \caption{\small Summary of theoretical predictions for the $a_0(1710)$ mass made before its first measurement in 2021~\cite{BaBar:2021fkz} (in MeV).}
\vspace{-0.5cm}
  \begin{center}
  \begin{tabular}{|c||c|c|c|c|c|c|c|}
    \hline
    \small Model & \small Relativized & \small Regge & \small Unitarized & \small Relativized & \small Extended & \small Regge & \small Unitarized \\
    \small  approach & \small quark model & \small analysis & \small amplitudes & \small quark model & \small lin. $\sigma$-model & \small analysis & \small amplitudes \\
    \small [ref.] (year) & \small \cite{Godfrey:1985xj} (1985) & \small \cite{cl2} (2007) & \small \cite{Geng:2008gx} (2009) & \small \cite{Ebert:2009ub} (2009) & \small \cite{Parganlija:2016yxq} (2017) & \small \cite{Wang:2017pxm} (2018) & \small \cite{Du:2018gyn} (2018) \\
    \hline
    &&&&&&&\\
    \small Prediction & $1780$ & $1700\pm60$ & $1777$ & $1679$ & $1790\pm35$ & $1774$ & $1770\pm20$
    \\
    \hline
  \end{tabular}
  \end{center}
\end{table}

As David Bugg noted in his review~\cite{bugg}, the masses of light non-strange mesons tend to cluster around certain mass values~\cite{bugg} (see Fig.~4 in that review). Later a detailed study of the
arising meson clusters showed that the averaged positions of the corresponding towers of states follow almost perfect linear Regge trajectory~\cite{cl1b} (independently it was observed also
in Refs.~\cite{klempt,shif}): A global fit of the data performed in~\cite{cl2,cl2b} lead to the following approximate relation for the averaged mass (in GeV$^2$),
\begin{equation}
\label{6}
M^2_{\text{exp}}(N)\approx1.14(N+0.54), \qquad N=0,1,2,3,4,
\end{equation}
where $N$ enumerates the clusters. The ground states below 1~GeV were excluded from the fit (i.e., the states corresponding to $N=0$ in~\eqref{6}).
Thus, for $M(N)$ we have (in MeV): $M(0)\approx 785$, $M(1)\approx 1325\pm90$, $M(2)\approx 1700\pm60$,
$M(3)\approx 2000\pm40$, $M(4)\approx 2270\pm 40$. The data used in~\cite{cl2b} for derivation of~\eqref{6} are displayed in Table~2.
The five question marks in Table~2 denote the predicted states. One of them is the $a_0$-meson with the predicted mass of $1700\pm60$~MeV.
This prediction was explicitly stated in Ref.~\cite{cl2}.

\begin{table}
\caption{\small The masses (in MeV) of light non-strange mesons used in the analysis of Ref.~\cite{cl2b}
(the data were taken from the Particle Data, the issue of year 2006, and compilation~\cite{bugg}).
The resulting averaged masses for each cluster $\overline{M}(N)$, $N=0,1,2,3,4$, are indicated in the bottom line (from Ref.~\cite{cl2}).
The question mark stands for the predicted states in the corresponding mass range.}
\begin{center}
\begin{tabular}{|llllll|}
\hline
Meson & $M(0)$ & $M(1)$ & $M(2)$ & $M(3)$ & $M(4)$\\
\hline
$\pi$&$135$&
$1300\pm100$&$1812\pm14$& $2070\pm35$& $2360\pm25$\\
$\eta$& 
&$1294\pm4$& $1760\pm11$& $2010^{+35}_{-60}$& $2285\pm20$\\
$\omega$& $782.65\pm0.12$&
$1400\div1450$& $1670\pm30$& $1960\pm25$&\!\!\!\!\!\!
\begin{tabular}{l}
$2205\pm30$\\
$2295\pm50$\\
\end{tabular}\\
$\rho$&$775.5\pm0.4$&
$1459\pm11$& $1720\pm20$& \!\!\!\!\!\!
\begin{tabular}{l}
$1900\pm?$\\
$2000\pm30$\\
\end{tabular}&\!\!\!\!\!\!
\begin{tabular}{l}
$2110\pm35$\\
$2265\pm40$\\
\end{tabular}\\
$f_0$&& 
$1200\div1500$& $1770\pm12$& $2020\pm38$& $2337\pm14$\\
$a_0$&& 
$1474\pm19$& \,\,\,\,\,\,\,\,\,\,\,{\large\bf ?} & $2025\pm30$& \,\,\,\,\,\,\,\,\,\,\,{\large\bf ?}\\
$a_1$&& $1230\pm40$& $1647\pm22$&$1930^{+30}_{-70}$&$2270^{+55}_{-40}$\\
$f_1$&&  $1281.8\pm0.6$& \,\,\,\,\,\,\,\,\,\,\,{\large\bf ?} & $1971\pm15$& $2310\pm60$\\
$h_1$&& $1170\pm20$& $1595\pm20$ & $1965\pm45$& $2215\pm40$\\
$b_1$&& $1229.5\pm3.2$& $1620\pm15$ & $1960\pm35$& $2240\pm35$\\
$f_2$&& $1275.4\pm1.1$& $1638\pm6$& \!\!\!\!\!\!
\begin{tabular}{l}
$1934\pm20$\\
$2001\pm10$\\
\end{tabular}&\!\!\!\!\!\!
\begin{tabular}{l}
$2240\pm15$\\
$2293\pm13$\\
\end{tabular}\\
$a_2$&& $1318.3\pm0.6$& $1732\pm16$&\!\!\!\!\!\!
\begin{tabular}{l}
$1950\pm40$\\
$2030\pm20$\\
\end{tabular}&\!\!\!\!\!\!
\begin{tabular}{l}
$2175\pm40$\\
$2255\pm20$\\
\end{tabular}\\
$\pi_2$&&& $1672.4\pm3.2$& $2005\pm15$& $2245\pm60$\\
$\eta_2$&&& $1617\pm5$& $2030\pm16$& $2267\pm14$\\
$\omega_3$&&& $1667\pm4$& $1945\pm20$& \!\!\!\!\!\!
\begin{tabular}{l}
$2255\pm15$\\
$2285\pm60$\\
\end{tabular}\!\!\!\!\!\!
\\
$\rho_3$&&& $1688.8\pm2.1$& $1982\pm14$&\!\!\!\!\!\!
\begin{tabular}{l}
$2300^{+50}_{-80}$\\
$2260\pm20$\\
\end{tabular}\\
$\omega_2$&&&\,\,\,\,\,\,\,\,\,\,\,{\large\bf ?}& $1975\pm20$& $2195\pm30$\\
$\rho_2$&&&\,\,\,\,\,\,\,\,\,\,\,{\large\bf ?}& $1940\pm40$& $2225\pm35$\\
$f_3$&&&& $2048\pm8$& $2303\pm15$\\
$a_3$&&&& $2031\pm12$& $2275\pm35$\\
$h_3$&&&& $2025\pm20$& $2275\pm25$\\
$b_3$&&&& $2032\pm12$& $2245\pm50$\\
$a_4$&&&& $2005^{+25}_{-45}$& $2255\pm40$\\
$f_4$&&&& $2018\pm6$& $2283\pm17$\\
$\omega_4$&&&&& $2250\pm30$\\
$\rho_4$&&&&& $2230\pm25$\\
$\pi_4$&&&&& $2250\pm15$\\
$\eta_4$&&&&& $2328\pm38$\\
$\omega_5$&&&&& $2250\pm70$\\
$\rho_5$&&&&& $2300\pm45$\\
\hline
&&&&&\\
$\overline{M}$ & $\approx780$ & $1325\pm90$ & $1700\pm60$ & $2000\pm40$ & $2270\pm 40$  \\
\hline
\end{tabular}
\end{center}
\end{table}

The observed pattern of approximate degeneracy has a nice physical interpretation proposed in Ref.~\cite{cl2b}.
Following~\cite{ani,bugg}, one can classify light non-strange mesons by their $(L,n)$ values, where $L$ and $n$ are
the orbital and radial quantum numbers, correspondingly. The emerging classification is displayed Table~3. As can be seen from Table~3,
the states with equal $N=L+n$ are approximately degenerate
(the corresponding boxes in Table~3 belong to the same diagonal formed by fixed $L+n$).

\begin{table}
\vspace{-1cm}
\caption{
\small Classification of light non-strange mesons according
to the values of $(L,n)$ from  Ref.~\cite{cl2b}.}
\begin{center}
\begin{tabular}{|c|c|c|c|c|c|}
\hline
\begin{tabular}{c}
\begin{picture}(15,15)
\put(0,12){\line(1,-1){15}}
\put(-2,-3){$L$}
\put(10,7){$n$}
\end{picture}\\
\end{tabular}
& 0 & 1 & 2 & 3 & 4 \\
\hline
0
&
\begin{tabular}{c}
$\pi(140)$\\
$\rho(770)$\\
$\omega(780)$\\
\end{tabular}
&
\begin{tabular}{c}
$\pi(1300)$\\
$\rho(1450)$ \\
$\omega(1420)$ \\
$\eta(1295)$\\
\end{tabular}
&
\begin{tabular}{c}
$\pi(1800)$\\
$\rho(?)$\\
$\omega(?)$\\
$\eta(1760)$\\
\end{tabular}
&
\begin{tabular}{c}
$\pi(2070)$\\
$\rho(1900)$ \\
$\omega(?)$\\
$\eta(2010)$\\
\end{tabular}
&
\begin{tabular}{c}
$\pi(2360)$\\
$\rho(2150)$\\
$\omega(2205)$ \\
$\eta(2285)$\\
\end{tabular}
\\
\hline
1
&
\begin{tabular}{c}
$f_0(1370)$\\
$a_0(1450)$ \\
$a_1(1260)$\\
$f_1(1285)$\\
$b_1(1230)$\\
$h_1(1170)$\\
$a_2(1320)$\\
$f_2(1275)$\\
\end{tabular}
&
\begin{tabular}{c}
$f_0(1770)$\\
$a_0(?)$\\
$a_1(1640)$\\
$f_1(?)$\\
$b_1(1620)$ \\
$h_1(1595)$ \\
$a_2(1680)$\\
$f_2(1640)$\\
\end{tabular}
&
\begin{tabular}{c}
$f_0(2020)$\\
$a_0(2025)$\\
$a_1(1930)$ \\
$f_1(1971)$\\
$b_1(1960)$\\
$h_1(1965)$\\
$a_2(1950)$ \\
$f_2(1934)$\\
\end{tabular}
&
\begin{tabular}{c}
$f_0(2337)$\\
$a_0(?)$\\
$a_1(2270)$ \\
$f_1(2310)$\\
$b_1(2240)$\\
$h_1(2215)$\\
$a_2(2175)$ \\
$f_2(2240)$\\
\end{tabular}
&\\
\hline
2
&
\begin{tabular}{c}
$\rho(1700)$\\
$\omega(1650)$\\
$\pi_2(1670)$\\
$\eta_2(1645)$\\
$\rho_2(?)$\\
$\omega_2(?)$\\
$\rho_3(1690)$\\
$\omega_3(1670)$\\
\end{tabular}
&
\begin{tabular}{c}
$\rho(2000)$\\
$\omega(1960)$\\
$\pi_2(2005)$\\
$\eta_2(2030)$\\
$\rho_2(1940)$\\
$\omega_2(1975)$\\
$\rho_3(1982)$\\
$\omega_3(1945)$\\
\end{tabular}
&
\begin{tabular}{c}
$\rho(2265)$\\
$\omega(2295)$ \\
$\pi_2(2245)$\\
$\eta_2(2267)$\\
$\rho_2(2225)$\\
$\omega_2(2195)$\\
$\rho_3(2300)$ \\
$\omega_3(2285)$\\
\end{tabular}
&  &\\
\hline
3
&
\begin{tabular}{c}
$f_2(2001)$\\
$a_2(2030)$\\
$f_3(2048)$\\
$a_3(2031)$\\
$b_3(2032)$\\
$h_3(2025)$\\
$f_4(2018)$\\
$a_4(2005)$\\
\end{tabular}
&
\begin{tabular}{c}
$f_2(2293)$\\
$a_2(2255)$\\
$f_3(2303)$\\
$a_3(2275)$\\
$b_3(2245)$\\
$h_3(2275)$\\
$f_4(2283)$\\
$a_4(2255)$\\
\end{tabular}
&  &  &\\
\hline
4
&
\begin{tabular}{c}
$\rho_3(2260)$\\
$\omega_3(2255)$\\
$\rho_4(2230)$\\
$\omega_4(2250)$ \\
$\pi_4(2250)$\\
$\eta_4(2328)$\\
$\rho_5(2300)$\\
$\omega_5(2250)$\\
\end{tabular}
&  &  &  &\\
\hline
\end{tabular}
\end{center}
\end{table}

The dependence of excitation energy on the sum $L+n$ is a characteristic feature of Coulomb spectrum. The resulting dynamic degeneracy is known to arise
due to the increased $O(4)$ symmetry of the Coulomb potential. For this reason, the resulting degeneracy is often called hydrogen-like. But in light mesons
the functional dependence of excitation energy on $L+n$ is different from the spectrum of hydrogen atom and has the approximate form
\begin{equation}
\label{1}
M^2(L,n) \approx a\left(L+n+\frac12\right),
\end{equation}
with the phenomenological slope $a\approx 1.14$~GeV$^2$. The physical origin of relation~\eqref{1} has a natural qualitative explanation within the
framework of semiclassical hadron string approach~\cite{cl2b,Afonin:2024egd}. Further development of this topic is given in Refs.~\cite{Afonin:2024egd,Bicudo:2007wt,Afonin:2007sv,Mezoir:2008vx,Afonin:2009zk,Bicudo:2009hm}.

The relation~\eqref{1} is supposed to describe, in the first approximation, the observed approximate degeneracy in light non-strange mesons. Tables 2 and 3, however, contain $\eta$-mesons
and some other resonances with dominant strange component in their decay products.
To avoid possible confusion, this issue should be clarified. The typical excitation energy in light mesons is 300--500~MeV, which is more than twice the mass of the s-quark.
From the viewpoint of quantum field theory, this means that a part of the excitation energy can exist in the form of $s\bar{s}$-pair. This effect appears to significantly blur
the distinction between strange and non-strange mesons in the highly excited part of their spectrum. Almost all highly excited light mesons have an admixture of $s\bar{s}$
component in their decay products which is materialized by the presence of $K\bar{K}$ or $\eta$ in final states. But the presence of $s\bar{s}$ component does not necessarily mean
that it was a part of wave function of a given meson --- $s\bar{s}$-pair may simply have formed out of the gluon field, thereby causing the strong decay of that meson.
Moreover, decays via such a formation of $s\bar{s}$-pair may predominate because the $s$-quark is much heavier than the $u$ and $d$, so it will be slower for a given energy, which is more favorable for the production of final stable hadrons. For example,  the letter "$\eta$" in the excited part of the spectrum, in reality, just denotes the isosinglet pseudoscalar particle --- the given notation does not mean that the resonance under consideration has the same quark composition as the ground state with these quantum numbers. The ground state in the $\eta$-channel was excluded in Tables 2 and 3 because the strange component is dominant in that meson but the same cannot be claimed for the high radial and orbital excitations of the $\eta$-meson (in particular, such excitations were observed in proton-antiproton annihilation~\cite{pdg,ani}, where there is almost no $s\bar{s}$ component in the initial state). Also were excuded light unflavored mesons for which the absolute dominance of $s\bar{s}$ component in their quark composition is well established, for example, $\phi$-mesons.

The given note was not completely taken into account in Refs.~\cite{cl2,cl2b}: In Tables 2 and 3, the scalar resonance $f_0(1710)$ is replaced by the much less established $f_0(1770)$
due to the bias that $f_0(1710)$ must represent a mostly $s\bar{s}$ state or a scalar glueball.

Table~3 contains eight predicted states. Three of them are identified with excited $S$-wave vector mesons which should be degenerate in mass with the corresponding known $D$-wave vector mesons according to~\eqref{1}. Most likely, these $S$- and $D$-wave states cannot be distinguished experimentally and we excluded these predictions in Table~2.
The predicted $\rho_2$ and $\omega_2$ are unlikely to be established --- although the quark model predicts such states, the formation of these resonances appears to be strongly suppressed,
and there are no well-established mesons with given quantum numbers. Thus we are left with three realistic predictions: $a_0(1700)$, $f_1(1700)$, and $a_0(2270)$, where the number in
brackets is the expected mass in MeV. It looks natural to identify the first one with the meson $a_0(1710)$ of the Particle Data~\cite{pdg}. The remaining two are waiting to be discovered.

A thorough statistical analysis of the data in Ref.~\cite{cl2,cl2b} was not performed. To enhance the rigor of made predictions, we did such a statistical analysis using data from~\cite{cl2,cl2b}.
The Shapiro-Wilk statistical test for normality showed that the data in each cluster (except $N=0$ where there is no enough data) follows normal distribution with
a high probability, exceeding 0.9 with confidence level 95\%. Fixing this confidence level, we obtained (in GeV$^2$)
\begin{equation}
\label{2}
\bar{M}^2(N) = (1.12\pm0.03)N+(0.61\pm0.08)\simeq (1.12\pm0.03)(N+0.54\pm0.09),
\end{equation}
that is close to the experimental fit~\eqref{6}. We have tried other exercises with statistics using this data but the changes we got were minor.

The spectrum of highly excited light mesons in the modern Particle Data~\cite{pdg} is partly different from the spectrum used in Table~2. A similar analysis with new data has been recently
performed in Ref.~\cite{Afonin:2024egd} under various assumptions. The hydrogen-like degeneracy was confirmed, the extracted parameters turned out to be close to those displayed in~\eqref{2}. This justifies {\it a posteriori} the prediction of $a_0(1710)$ discussed above. Since meson states lying on Regge trajectories and daughter Regge trajectories (radial excitations)
represent ordinary quark-antiquark states, the resonances $a_0(1710)$ and $f_0(1710)$ should be conventional mesons.

\end{document}